\documentclass[12pt,preprint]{aastex}

\newcommand{\jmh}{{\em J-H}}
\newcommand{\jmk}{{\em J-K}}
\newcommand{\hmk}{{\em H-K}}

\newcommand{\isotwothree}{[6.7\mu{\rm m}]-[14\mu{\rm m}]}

\def \vs          {{\it vs.} }
\def \etal        {{\it et~al.} }
\def \eg          {{\it e.g.} }

\shorttitle{ISO observations of L 1641-N}
\shortauthors{Ali \& Noriega-Crespo}

\begin{document}

\title{A Mid-IR survey of L 1641-N with ISOCAM}

\author{Babar Ali}

\affil{Infrared Processing and Analysis Center, MC 100-22\\
California Institute of Technology, Pasadena, CA 91125}

\email{babar@ipac.caltech.edu}

\and

\author{Alberto Noriega-Crespo}

\affil{Spitzer Science Center / Infrared Processing and Analysis Center,
MC 220-6\\ California Institute of Technology, Pasadena, CA 91125}

\email{alberto@ipac.caltech.edu}

\begin{abstract}
We present results from our analysis of the Infrared Space Observatory (ISO) data on the L~1641-N outflow region located in the Orion~A molecular cloud.  The data were obtained with the  array camera (ISOCAM) using two broad-band filters, LW2 (6.7~\micron), \& LW3 (14~\micron), and the narrow-band circular variable filter (CVF) which provides low resolution (R=$\lambda/\delta\lambda\sim40$) spectra in the 5--17~\micron\ region.
\par
We detect a total of 34 sources in the $7.65^{\prime}$~x~$8.40^{\prime}$\ region covered by the
broad-band filters.  Four of these sources have no reported detection in previous studies of the region.  
The CVF data are available for only the central $3.2^{\prime}$~x~$3.2^{\prime}$\ portion of the total region, providing spectra for the brightest 7 of the sources in that region.
\par
We find that the source previously identified as the near-IR counter-part to the IRAS detected point-source (IRAS 05338-0624) is not the brightest source in the wavelength region of the IRAS 12~\micron\ filter.  We find instead that a nearby object (within the beam of IRAS and not detected at near-IR wavelengths) outshines all others sources in the area by a factor of $\sim$2.  We submit that this source is likely to be the IRAS detected point source. 
\par
A comparison of the near-IR (\jmh\ \vs\hmk) and mid-IR (\hmk\ \vs $\isotwothree$) color-color plots shows that while at near-IR wavelengths only four (4) of the sources show evidence for emission above the values predicted from photospheric emission alone (hereafter referred to as excess emission), at least 85\% of all sources show evidence for excess emission at mid-IR wavelengths.  This result supports similar conclusions from  $L$-band surveys.
\par
The CVF spectra suggest a range of evolutionary status in the program stars ranging from embedded YSOs to young disks.  When combined with optical and near-IR age estimates, these results show active current star-formation in the region that has been on-going for at least 2 Myr.
\end{abstract}

\section{Introduction} \label{sec-intro}
\par
Intensive observational and theoretical studies of young, optically obscured star-forming regions over the last four decades suggest that stars like our Sun begin their lives in the midst of cold and dense molecular clouds. During the earliest phases of their evolution, these young stars are embedded behind 50-200 magnitudes of visual extinction , and likely develop extended bipolar outflows and jets in the early phases of their evolution towards the main-sequence (for a review, see Lada \& Lada 2002).   The Orion giant molecular cloud complex is the closest example of such a star-forming region that offers both massive and low-mass stellar population.
\par
The L~1641-N region is located in the southern part of the Orion~A molecular cloud. Attention was especially drawn towards L~1641-N by IRAS with the detection of a bright point source (IRAS 05338-0624). Subsequently, \citet{fukui} reported discovery of two well-separated outflow lobes surrounding the IRAS source.  Further optical and near-infrared imaging studies revealed an enhancement of stellar density near the IRAS source at near-IR wavelengths (Chen \etal\ 1996; Hodapp \etal\ 1993), located "among the highest concentrations of HH [Herbig-Haro] objects known anywhere in the sky"  (Reipurth \etal 1998).
\par
The presence of the HH outflows implies that the L~1641-N region is one of the youngest star-forming environment in the Galaxy.  Dynamic age estimates of the outflow sources presently observed in the Galaxy constrain outflow ages to less than $10^5$~years (Fukui
1989; Fukui \etal 1993).  Studying these young, optically obscured environments at mid-infrared wavelengths offers several advantages: (i) stars in the earliest phases of evolution are heavily obscured and are frequently undetected at even near-IR wavelengths.  (ii) With imaging data alone, it is difficult to distinguish between actual photospheric emission and reprocessed emission (reflected light).  Thus, near-IR reflection nebulae are often confused with actual sources.  (iii) At mid-IR wavelengths, both the stellar photosphere and the colder star-forming environment are visible.
\par
However, the mid-IR region is difficult to observe from the ground because of the presence of high background from the warm Earth atmosphere. The Infrared Space Observatory (ISO, Kessler \etal\ 1996) provided the first space-based observatory with the ability to carry out sensitive observations at mid-IR wavelengths.  The ISO observed spectra of the embedded Young Stellar Objects (YSOs) revealed, for the first time, the presence of silicate, CO, water (vapor and ice) absorption features in the photospheric continuum of the stars (see for example, Gibb \etal~2000).  These features are thought to arise in the colder star-forming environment in which the stars are forming.  Thus, the mid-IR region provides information both on the star itself and about the physical conditions in which they form.
\par
In this contribution we provide an analysis of ISO CAMera (ISOCAM, Cesarsky \etal~1996) mid-infrared broad-band and low-resolution Circular Variable Filter (CVF) observations of the L~1641-N region.  Section~\ref{sec-obs} describes these observations in detail.  Section~\ref{sec-phot} describes our source detection, photometry and cross-correlation with other optical and near-IR sources.  We present  and discuss our results in Section~\ref{sec-results}.  Finally, the conclusions are summarized in Section~\ref{sec-summary}.

\section{Observations and data reduction} \label{sec-obs}
\par
All data for this contribution were obtained from the ISO Archive (Salama \etal~2002).  The observing parameters and principle investigators for the observations used in this contribution are summarized in Table~\ref{tbl-obsum}.  Column~1 in Table~\ref{tbl-obsum} lists the unique ISO identifier, the TDT number, column 2 lists the ISO Astronomical Observing Template (AOT) used to obtain the data, columns 3 and 4 identify the filters and the associated wavelengths.  For the LW2 and the LW3 broad-band filters, the central wavelength is listed, and for the Circular Variable Filter (CVF) the wavelength coverage is listed. Column 5 lists the total exposure time and column 6 lists the principle investigator (PI) for the observing proposal.
\par
The data were reduced using the CAM Interactive Analysis (CIA, Ott \etal\ 2001) software.  The raw data files (CISP \& IIPH files) were obtained from the ISO Archive and reprocessed locally.  The data reduction comprised both ``standard'' CIA and customized reduction modules.  The reduction procedure -- in order of steps performed -- is listed below.
\par\noindent
{\bf{\it Dark Current Correction}} Over the lifetime of ISO, the ISOCAM dark frames showed cyclical variations in the observed levels of dark current.  The periodicity, level and causes of this ISOCAM dark behaviour are described in \citet{camhandbook}.  The CIA routine 'corr\_dark' was used with the 'VILSPA' option to subtract the dark current from each individual frame.  The 'VILSPA' dark model accounts for all known ISOCAM dark current variability modes.
\par\noindent
{\bf{\it Cosmic Ray Removal}} The CIA routine 'deglitch' with the multi-resolution median filtering option was used to provide an initial assessment of the cosmic ray hits.  This automatic procedure identified most cosmic ray events; however, a small ($<2\%$) of the pixels were incorrectly identified as cosmic ray hits.  Thus, the results from the automated procedure were augmented by visual inspection of the data cube.
\par\noindent
{\bf{\it Non-linearity corrections}} The ISOCAM detectors show the so-called ``transient'' behaviour -- a non-linear response to input flux.  The theory of detector transients is described by, for example, \citet{haegel}.  The ISOCAM transients are reviewed by \citet{coulais}.  We used the 'Fouks-Shubert' removal within CIA to correct for transients.  This correction did not remove all transient artifacts in the broad-band rasters (where the effects are the largest).  We simply masked out the uncorrected data and replaced these values by redundant values from the overlapping raster observation.
\par\noindent
{\bf{\it Flat-field correction}} A single flat field for the entire array was constructed by median-combining all raster segments.  This option is preferable to using a library flat-field, since a custom corrected flat-field is not affected by the ``jitter'' in the ISOCAM optics, which moves the optical flat-field pattern on the ISOCAM array.
\par
The CIA routines 'raster\_scan' was used to assemble the final mosaic of the broad-band data.

\section{Source detection and  photometry}
\label{sec-phot}
\par
All sources were identified using the DAOPHOT \citep{daophot} \textit{find} algorithm as implemented within the Interactive Data Language (IDL) environment.  Visual inspection of the detected sources revealed that most DAOPHOT detections were either bright knots in the nebulosity or ISOCAM artifacts.  The final sample of stars is, hence, drawn by confirming each detection with visual inspection or by comparison with the 2MASS catalog. Given the unreliable nature of source finding algorithm used -- complicated by varying nebulosity and undersampled images, and uncertain PSF -- there may be additional sources but these cannot be confidently seperated from non-sources.  Hence, only those which are verified by one of the above methods are included in our final list.
\par
We used a 3 pixel by 3 pixel square aperture for photometry on all sources.  We use the ISOCAM PSF library (available within CIA) to correct for flux outside of our adopted square aperture.  This correction is 12\% and 15\% for LW2 and LW3 filters, respectively.  We use a global background sky estimate for all sources.  To obtain the sky/background per pixel, we determine the median value in all regions that show no  extended emission.  Similar apertures were also used to extract the CVF spectra of the sources, where CVF data is available.  No background correction is applied to the CVF spectra.
\par
We used the CIA's internal flux calibration routine 'adutomjansky' to convert source counts to physical flux units.  This calibration is based on standard stars and is discussed by \cite{camhandbook} and gives the following values for the LW2 and the LW3 filters:
\par\noindent
LW2: 1 mJy = 2.3303 counts \\
LW3: 1 mJy = 1.9733 counts
\par
We refined the initial astrometry estimates for the sources (based on ISO's pointing information) by comparing the images to 2MASS images and using the 2MASS coordinates, and by applying the optical distortion correction given by \citet{distortion-paper}.  Given the low number and uncrowded (source PSFs are well-separated) field, we cross-identified our ISO sources visually with 2MASS catalog and with the near-IR study of \citet{chen}.
\par
The final photometry and astrometry measurements for all sources discussed in this contribution is listed in Table~\ref{tbl-phot}. Column~1 lists our assigned identification (ID) number.  Column~2 list \citet{chen} designations for the sources.  We list multiple cross-identified sources when a single unambiguous determination could not be made.   Columns~3~\&~4 list the 2MASS coordinates for the source.  Sources with ID numbers 10, 18, 19, \& 20 have no identified 2MASS counterpart.  For these sources the ISO coordinates are listed.  To ensure uniformity, we computed coordinate offsets between ISO and 2MASS positions.  For sources without 2MASS detections the listed ISO coordinates are corrected for the aforementioned positional offsets.  The magnitude of the positional offset is 0.24 and 7.7 arc-seconds in right-ascension and declination, respectively.  Columns~5 and 6 list the ISO flux in milli-Jansky.  The errors are computed from the standard deviation observed among the multiple measurements of the object.  For eight of the sources, only one measurement is available.  We assigned 20\% errors to these sources, which is the average error percentage for all sources with multiple measurements.  These stars are identified in Table~\ref{tbl-phot}.  Four objects were discernible only in the final mosaicked image of the region.  For these objects, we used the XPHOT photometry package available within CIA (Ott \etal\ 2001) to obtain aperture photometry.  These sources are also identified in Table~\ref{tbl-phot}.  Columns 7-9 list the near-infrared flux measurements from 2MASS. 

We compared our photometry with that made available by the ISOCAM pipe-line (OLP version 10.0, Blommaert \etal 2003).  We consider the OLP photometry to be less reliable in that several of the artifacts (\eg\ residual images from transients, Section~\ref{sec-obs}) are visible in the OLP reduced data.  These were removed from our reduction.  Our comparison shows that mean differences between our photometry and OLP photometry to be 18\% and 34\% for the LW2 and the LW3 filters, respectively.

\section{Results and discussion} \label{sec-results}
\par
We detect a total of 34 sources in the region covered by the broad-band filters.  The spatial coverage and sensitivity of the narrow-band filter data allowed us to obtain low-resolution spectra of only 7 of these sources.  All but 4 of the sources are previously reported by the Two Micron All Sky Survey (2MASS) and/or  the near-IR survey of  \citet{chen}.  Two of the sources (\#10 and \#18, in our numbering scheme) are only detected at mid-IR wavelengths: by ISOCAM and by \citet{chen} in their M-band images.  These results are further discussed below.

\subsection{The counterpart to IRAS 05338-0624} \label{sec-irascounterpart}
\par
\citet{chen} suggest source \#18, in our numbering scheme, as the likely near-IR counterpart to the IRAS detected point source based on their {\it M}-band imaging survey.  Source \#18 is within 2 arc-seconds of the IRAS published position and is the brightest source at {\it M}-band in the \citet{chen} survey.   \citet{stanke} also detected a bright (0.4 Jy) point source at 10\micron\ coincident with the location of source \#18.  However, \citet{stanke} argue that IRAS 05338-0624 cannot be unambiguously associated with source \#18 because a 2.7-mm dust continuum source (Chen \etal\ 1993, 1995) is also coincident with the position uncertainty ellipse of the IRAS source (Figure ~\ref{fig-m2}, also see Figure 4 of Stanke \etal.\ 1997).

In the ISO image, source \#10 outshines source \#18 by approximately a factor of two.  Source \#10 is outside the field-of-view of the \citet{chen} {\it M}-band survey and the \citet{stanke} 10\micron\ image, and is $\sim$1 arc-minute away from the IRAS source position.  There are no other comparably bright sources detected in the ISO images.   The narrow-band CVF spectra (Figure~\ref{fig-cvfspec}) include the wavelength range of the shortest IRAS filter and are useful for obtaining a more appropriate flux comparison with IRAS measurements.  We used the ISO Spectral Analysis Tool (ISAP) and the IRAS filter transmission curves to simulate IRAS photometry for sources \#18 and \#10 using the CVF spectra.  We find source fluxes of 284 and 481 mJy (30\% errors) for sources \#18, and \#10, respectively.  The IRAS published photometry for IRAS 05338-0624 is 481 mJy.

Despite the excellent agreement between the fluxes of IRAS 05338-0624 and source \#10, we conclude that source \#18 remains the most likely counter part to IRAS 05338-0624 because it is the brightest source that is consistent with the position uncertainty ellipse of the IRAS source.   We rule out source \#10 because for it to be the IRAS counterpoint implies a positional error of $\sim$1 arc-minute in the IRAS position.  We consider this to be unlikely given that IRAS positions have been shown to be reliable to better than 7 arc-seconds (1-sigma) based on statistical comparison between IRAS and the Smithsonian Astrophysical Observatory (SAO) catalog point sources (cf. Section VII of the IRAS Explanatory Supplement, Beichman \etal 1988).  The 2.7-mm dust continuum source lacks a bright mid-IR counter part in the ISO image implying either that it is not a strong emitter at those wavelengths or that it has dimmed significantly in the decade between IRAS and ISO measurements.   

However, we also consider it unlikely that IRAS would not have spatially resolved and detected source \#10, which is the  brightest source in the ISO image and about 1 IRAS beam-width away from IRAS 05338-0624 (source \#18).  If we rule out  positional error in IRAS coordinates, then this implies that source \#10 must have brightened by about at least 200 mJy  (using point source detection limit of 0.3 Jy for IRAS) between the ISO and IRAS epochs.

\par
We also note that the proposed IRAS counterpart, source \#18, shows a broader PSF than other point sources in the area.  This is likely due to the presence of extended emission or emission knots which are unresolved by ISOCAM and seen in reflected light in the K-band as noted by \citet{chen}.  

\subsection{The nature of point-like sources}
\par
Figure~\ref{fig-nirclrs} shows the \jmh\ \vs \hmk, color-color relationship for the 30 sources for which near-IR photometry is available.  In constructing this relationship, we used the photometry from 2MASS database.   The solid lines in the figure show the colors of ordinary dwarf and giant stars from the compilation of \citet{bessell}.  The dotted lines in Figure~\ref{fig-nirclrs}  show the effect of extinction on the colors of the reddest giant and dwarf stars.  The distribution of the L1641-N stars on Figure~\ref{fig-nirclrs} is consistent with those from deeply embedded young stellar clusters. And, four stars, \#6, \#11, \#17, and \#21 show evidence for the so-called "excess' emission -- colors that are redder than predicted by simple foreground extinction.  This "excess" emission is though to arise from warm circumstellar material (see \eg\ Lada \& Lada 2003).  The emission itself is from warm dust in the circumstellar material.
\par
We investigated a similar color-color relationship by using the ISO photometry for which all 4 photometry measurements are available (24 stars).  The results are shown in Figure~\ref{fig-midclrs}: \jmk\ \vs $\isotwothree$.  As in Figure~\ref{fig-nirclrs}, the dotted line shown in Figure~\ref{fig-midclrs} shows the effect of foreground extinction, but in this case, for a star with intrinsic $\isotwothree$\ color of 0 mag.  At mid-IR wavelengths only a handful of estimates are available for the intrinsic colors of ordinary or standard stars.  These estimates show intrinsic values that are either 0 or close to 0 mag.  Thus, we adopted 0 mag as the typical $\isotwothree$\ color.  While 4 stars show evidence for excess emission in Figure~\ref{fig-nirclrs}, all but four, sources \#5, \#31, \#22, and \#34 show evidence for excess emission at mid-IR wavelengths. This can be explained simply as warm dust emits more at longer than near-IR wavelengths; thus it is easier to detect excess emission at mid-IR wavelengths than near-IR color-color plot.  A similar argument was presented for the Chamaeleon I population by \citet{nordh} and \citet{persi}.  Adding source \#10 and \#18 in this sample, we conclude that at least 85\% of all sources detected in the ISO LW2 filter show evidence of excess emission at mid-IR wavelengths.
\par
Given the presence of excess emission in most of the observed sources, we conclude that most sources are most likely physically associated.  A conclusion also reached by \citet{chen} in their analysis, who labeled it as a "stellar density enhancement".  The ISO narrow-band CVF spectra for the seven sources detected in the CVF filter shed further light on the nature and evolutionary status of the stars in this grouping.  These ISO spectra are shown in Figure~\ref{fig-cvfspec}.  Each panel of Figure~\ref{fig-cvfspec} shows the spectrum from the source identified in the left-hand-side of the panel.  Some of the prominent features associated with young stars are also labeled.
\par
Five of the 7 detected sources, \#11, \#13,  \#10,  \#18, and \#19, show the broad feature at 9.5 \micron, which is characteristic of silicate absorption.  Sources \#10, and \#19 further show features seen in the very youngest proto-stars such as W 33A (Gibb \etal\  2000).  These features are labeled in Figure~\ref{fig-cvfspec} and have been attributed to ices (CO$_2$, H$_2$O, etc.) in the environment of the young embedded stars (Gibb \etal\ 2000).  Source \#14 shows marginal detections at wavelengths consistent with H$_2$\  emission.  Sources \#11 and \#15, which show evidence for excess emission at near-IR wavelengths (see Figure~\ref{fig-nirclrs}), also show evidence for an emission feature at wavelengths consistent with the PAH emission at 11.2~\micron.  We rule out silicate emission because its peak is near 10~\micron.   PAH emission has been observed in some T-Tauri stars and roughly 50\% of the Herbig Ae/Be stars in the sample of \citet{meeus}.   PAH emission can signify the presence of tiny grains and (likely) UV radiation to excite the PAH.  This is consistent with the observed location of sources \#11 and \#15 on Figure~\ref{fig-nirclrs}.  Source \#19, the faintest object detected in the CVF spectra, has relatively poor signal-to-noise ratio and most narrow features seen in its spectra are likely instrumental artifacts.  The only significant detection in the spectra of source \#19 is the broad silicate feature.
\par
\citet{evans} provide a  crude evolutionary sequence for the youngest stars based on spectral features in the mid-infrared wavelengths (see their Figure 8).  We use this sequence to interpret our CVF spectra.  Sources \#10, \#18, and possibly \#19 are, thus, in the earliest phase of evolution, labeled as 'embedded YSOs' by \citet{evans}.  The ice features in the spectra of sources \#10,  \#18 are seen in the earliest phases when the proto-star is deeply embedded in a circumstellar envelope.  As the circumstellar environment matures, the ices give way to molecular and atomic features during the 'Emerging YSO' phase.  Source \#14 in our survey is consistent with such an object.  Sources \#11, \#15 show PAH emission but otherwise a featureless spectrum.  These are consistent with the 'Young disk' phase of \citet{evans}.  The evolutionary status of source \#13 is unclear.  The presence of silicate absorption is indicative of the embedded YSO phase but the lack of features from other ices is puzzling.  Perhaps this is an intermediate object between the Embedded and Emerging YSO phase.
\par
In our interpretation of the spectra the ages of stars are of the order $\sim$few$\times10^4$~years for the embedded YSOs to $\sim$few$\times10^5$~years for stars with 'Young disk'.  These ages are consistent with \citet{fukui} based on their dynamical estimates using the outflows.  However,  \citet{chen} estimate the cluster population to be 1-2 Myr based on 3 stars detected by optical and near-IR surveys.  This inconsistency is likely just the difference between the optically detected and the mid-IR detected population of L 1641-N.  The combination of the two estimates establish this region as one of the youngest regions in Orion with active star-formation and age spread of order 2~Myr among the stars.

\section{Summary} \label{sec-summary}
Our analysis of the ISOCAM broad-band and CVF data produced the following main results:
\par
We detect a total of 34 sources in the region covered by the broad-band filters.  The CVF spectra are available for  7 of these sources.  Two sources (\#10 and \#18) are only detected at mid-IR wavelengths. 
\par
We find that the source previously identified as the counter-part to the IRAS detected point-source in the region is not the brightest source at the IRAS 12~\micron\ filter wavelengths.  We propose instead that the IRAS source is likely to be combination of sources with \#10 (in our listing) being the dominant one.  This source outshines all others by roughly a factor of 2.
\par
However, source \#18, previously associated with the IRAS source, shows a larger PSF than other point sources in the area.  This is likely due to the presence of extended emission or emission knots which are unresolved by ISOCAM.  We conclude that source \#18 remains the most plausible source of the outflows as suggested by \citet{chen}.
\par
A comparison of the near-IR (\jmh\ \vs\ \hmk) and mid-IR (\hmk\ \vs\ $\isotwothree$) color-color plots shows that while only two of the sources show evidence for ``excess'' emission, at least 85\% of all sources show evidence for excess emission at mid-IR wavelengths.  For two of these sources, CVF spectra are available and show emission features from PAH.  The presence of excess emission in most sources supports the assertion by \citet{chen} that these objects are physically associated, or at least co-located.
\par
The narrow-band CVF spectra of the 7 sources detected in the CVF show a range of evolutionary status for the stars, ranging from the earliest 'Embedded YSOs' to the 'Young disks'.  When combined with optical and near-IR age estimates, these results show active current star-formation in the region that has been on-going for 2~Myr.
\par
The observed absorption features in the spectra of the sources deserve further study in the form of detailed  modeling.  Future work will include full radiative transfer modeling to interpret the physical conditions responsible for the observed spectral features.  We plan to provide these results in a future contribution.

\acknowledgments{We thank the anonymous referee for helping us improve this manuscript.  A.N-C. research is partially supported by NASA through a contract with Jet Propulsion Laboratory, California Institute of Technology, and ADP Grant NRA0001-ADP-096. }

\clearpage


\begin{deluxetable}{lllccl}
\tablewidth{0pt}
\tablecaption{Summary of ISOCAM observations.\label{tbl-obsum}}
\tablehead{
\colhead{TDT} & 
\colhead{AOT} & 
\colhead{Filter} & 
\colhead{$\lambda$} & 
\colhead{Exposure Time} &
\colhead{PI} \\
\colhead{} &
\colhead{} &
\colhead{} & 
\colhead{(\micron)} &
\colhead{(sec)} &
\colhead{}
}
\startdata
69501706 & CAM01 & LW2 & 6.75       & 6916 & L. Nordh \\
         &       & LW3 & 15.0       &      & L. Nordh \\
82702009 & CAM01 & CVF & 5.4 - 16.8 & 60   & P. Andre \\
82702010 & CAM04 & CVF & 5.0 - 16.8 & 3607 & P. Andre \\
82702011 & CAM01 & CVF & 5.0 - 5.14 & 76   & P. Andre
\enddata
\end{deluxetable}

\clearpage
\thispagestyle{empty}

\begin{deluxetable}{ccccrrrrr}
\rotate
\tablewidth{0pt}
\tablecaption{Photometry of L 1641-N sources.\label{tbl-phot}}
\tablehead{
\colhead{ID} & 
\colhead{Other} & 
\colhead{$\alpha$} &
\colhead{$\delta$} &
\colhead{[6.7\micron]} &
\colhead{[14\micron]} &
\colhead{$J$} &
\colhead{$H$} &
\colhead{$K$} \\
\colhead{ } &
\colhead{Designation} &
\colhead{ } &
\colhead{ } &
\colhead{(mJy)} &
\colhead{(mJy)} &
\colhead{(mag)} &
\colhead{(mag)} &
\colhead{(mag)}
}
\startdata
01 & \nodata & 05 36 32.39 & -06 19 19.70 & 21.54$\pm$1.00 & 15.10$\pm$3.05 & 11.059$\pm$0.024 & 10.349$\pm$0.023 & 9.860$\pm$0.027 \\ 
02 & \nodata & 05 36 30.10 & -06 23  9.95  & 27.71$\pm$0.41 & 29.05$\pm$8.73 & 11.935$\pm$0.029 & 10.861$\pm$0.027 & 10.157$\pm$0.030 \\ 
03 & N28      & 05 36 27.72 & -06 23 12.16 & 12.79$\pm$1.28 & 8.74$\pm$1.69\tablenotemark{a} & 11.920$\pm$0.030 & 11.232$\pm$0.026 & 10.804$\pm$0.030 \\ 
04 & \nodata & 05 36 25.87 & -06 24 58.52 & 82.82$\pm$6.90 & 125.19$\pm$6.07 & 16.410$\pm$0.140 & 12.861$\pm$0.035 & 10.813$\pm$0.033 \\ 
05 & \nodata & 05 36 24.02 & -06 25 27.17 & 8.83$\pm$0.84 & 2.21$\pm$0.93 & 13.876$\pm$0.030 & 11.686$\pm$0.025 & 10.520$\pm$0.032 \\ 
06 & \nodata & 05 36 23.60 & -06 24 51.34 & 90.99$\pm$18.96 & 107.38$\pm$19.88 & 17.787$\pm$0.000 & 15.528$\pm$0.000 & 13.236$\pm$0.068 \\ 
07 & \nodata & 05 36 21.86 & -06 26  1.83  & 43.10$\pm$5.44 & 19.55$\pm$6.37 & 12.978$\pm$0.029 & 11.547$\pm$0.026 & 10.651$\pm$0.032 \\ 
08 & \nodata & 05 36 19.39 & -06 25 51.21 & 7.71$\pm$1.28 & 6.91$\pm$1.33\tablenotemark{a} & 13.009$\pm$0.026 & 12.278$\pm$0.024 & 11.762$\pm$0.030 \\ 
09 & N17       & 05 36 24.48 & -06 22 23.37 & 6.10$\pm$1.18 & 5.21$\pm$1.04\tablenotemark{a}\tablenotemark{b} & 12.707$\pm$0.025 & 11.552$\pm$0.027 & 11.049$\pm$0.032 \\ 
10 & \nodata & 05 36 24.58 & -06 22 41.33 & 205.22$\pm$2.05 & 436.96$\pm$44.95 & \nodata & \nodata & \nodata \\ 
11 & N29       & 05 36 23.79 & -06 23 11.27 & 75.60$\pm$2.67 & 76.42$\pm$3.23 & 14.162$\pm$0.054 & 12.357$\pm$0.040 & 10.805$\pm$0.034 \\ 
12 & N33      & 05 36 22.47 & -06 23 44.75 & 8.19$\pm$3.10 & 6.30$\pm$1.26\tablenotemark{a}\tablenotemark{b} & 13.986$\pm$0.030 & 12.560$\pm$0.028 & 11.694$\pm$0.033 \\ 
13 & N31-32 & 05 36 21.88 & -06 23 29.93 & 107.07$\pm$21.70 & 169.51$\pm$11.58 & 13.456$\pm$0.028 & 11.552$\pm$0.031 & 10.419$\pm$0.032 \\ 
14 & N30       & 05 36 20.50 & -06 23 22.14 & 6.62$\pm$0.84 & 6.64$\pm$1.95 & 15.739$\pm$0.084 & 13.290$\pm$0.034 & 11.897$\pm$0.033 \\ 
15 & N26      & 05 36 21.57 & -06 22 52.52 & 44.81$\pm$1.28 & 58.54$\pm$11.29\tablenotemark{a} & 14.028$\pm$0.031 & 12.068$\pm$0.028 & 10.957$\pm$0.032 \\ 
16 & N7,10   & 05 36 21.06 & -06 21 53.50 & 30.66$\pm$1.62 & 26.30$\pm$1.68 & 13.487$\pm$0.033 & 11.197$\pm$0.030 & 10.122$\pm$0.033 \\ 
17 & N5        & 05 36 22.64 & -06 21 27.71 & 4.22$\pm$1.28 & 3.63$\pm$0.73\tablenotemark{a}\tablenotemark{b} & 18.294$\pm$0.000 & 15.892$\pm$0.224 & 13.439$\pm$0.055 \\ 
18 & N1         & 05 36 19.50 & -06 22 12.71 & 184.22$\pm$57.75 & 169.78$\pm$10.38 & \nodata & \nodata & \nodata \\ 
19 & \nodata & 05 36 37.08 & -06 18 52.97 & 16.79$\pm$2.20 & 102.95$\pm$2.00 & \nodata & \nodata & \nodata \\ 
20 & \nodata & 05 36 18.94 & -06 22 56.67 & 9.15$\pm$1.28 & 2.35$\pm$0.62 & \nodata & \nodata & \nodata \\ 
21 & \nodata & 05 36 17.64 & -06 22 49.39 & 2.73$\pm$0.61 & 2.94$\pm$0.02 & 16.932$\pm$0.000 & 16.164$\pm$0.288 & 14.611$\pm$0.126 \\ 
22 & N34       & 05 36 15.23 & -06 23 55.70 & 11.76$\pm$0.48 & 0.91$\pm$0.65 & 9.839$\pm$0.024 & 9.504$\pm$0.026 & 9.473$\pm$0.031 \\ 
23 & N2         & 05 36 18.48 & -06 20 38.84 & 10.05$\pm$0.95 & 8.54$\pm$0.71 & 12.261$\pm$0.025 & 11.455$\pm$0.027 & 11.073$\pm$0.032 \\ 
24 & \nodata & 05 36 14.48 & -06 21  5.50  & 2.92$\pm$0.50 & \nodata & 15.806$\pm$0.081 & 13.106$\pm$0.033 & 11.855$\pm$0.034 \\ 
25 & \nodata & 05 36 12.97 & -06 23 33.06 & 17.82$\pm$0.67 & 11.57$\pm$0.50 & 11.596$\pm$0.030 & 10.403$\pm$0.032 & 9.935$\pm$0.035 \\ 
26 & \nodata & 05 36 12.61 & -06 23 39.58 & 14.85$\pm$2.20 & 5.16$\pm$1.03\tablenotemark{a}\tablenotemark{b} & 13.392$\pm$0.032 & 12.388$\pm$0.032 & 11.832$\pm$0.035 \\ 
27 & \nodata & 05 36 11.46 & -06 22 22.21 & 30.63$\pm$0.94 & 22.76$\pm$0.06 & 12.701$\pm$0.032 & 11.136$\pm$0.033 & 10.464$\pm$0.036 \\ 
28 & \nodata & 05 36 10.44 & -06 20  1.65  & 2.02$\pm$0.09 & \nodata & 11.691$\pm$0.024 & 11.001$\pm$0.026 & 10.898$\pm$0.031 \\ 
29 & \nodata & 05 36  8.34 & -06 24 37.84  & 3.88$\pm$1.28 & 2.27$\pm$0.44\tablenotemark{a} & 12.715$\pm$0.026 & 12.104$\pm$0.024 & 11.724$\pm$0.030 \\ 
30 & \nodata & 05 36  9.50 & -06 18 36.52  & 3.02$\pm$1.28 & \nodata & 11.801$\pm$0.025 & 11.038$\pm$0.026 & 10.900$\pm$0.030 \\ 
31 & \nodata & 05 36  6.86 & -06 23 35.31  & 7.18$\pm$1.45 & 2.00$\pm$0.94 & 10.027$\pm$0.024 & 9.785$\pm$0.028 & 9.737$\pm$0.030 \\ 
32 & \nodata  & 05 36  5.16 & -06 25 25.38  & 3.79$\pm$0.14 & \nodata & 11.936$\pm$0.025 & 11.256$\pm$0.022 & 10.962$\pm$0.029 \\ 
33 & \nodata & 05 36  6.94 & -06 18 53.53  & 6.27$\pm$0.87 & \nodata & 11.506$\pm$0.024 & 10.765$\pm$0.026 & 10.521$\pm$0.031 \\ 
34 & \nodata & 05 36  6.05 & -06 19 39.04  & 15.65$\pm$1.28 & 2.74$\pm$0.87 & 8.986$\pm$0.023 & 8.682$\pm$0.032 & 8.640$\pm$0.031
\enddata
\tablenotetext{a}{Assigned 20\% error.  See text for more detail.}
\tablenotetext{b}{Photometry obtained from the final mosaic}
\end{deluxetable}

\clearpage


\clearpage


\begin{figure}
\plotone{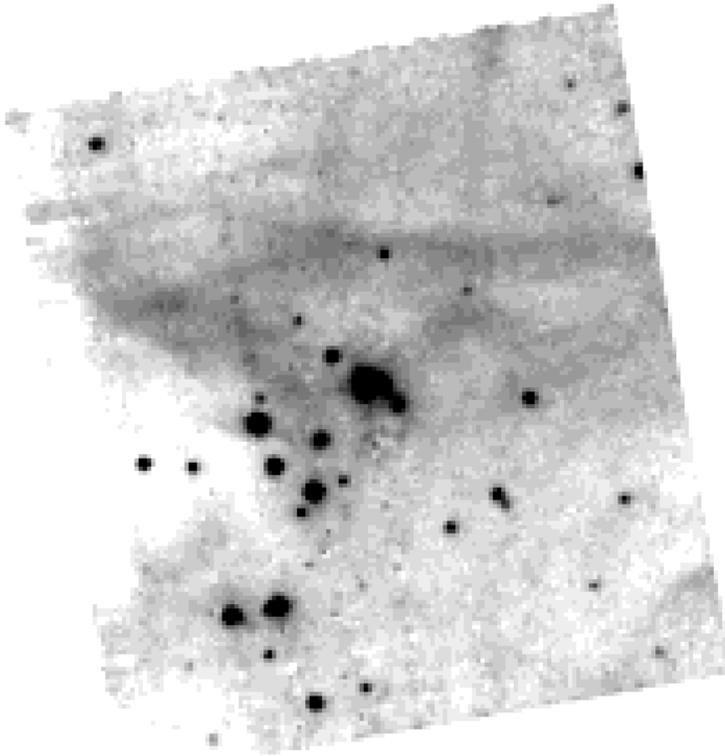}\caption{The final reduced ISOCAM mosaic of the L 1641-N region.  The image has been rotated such that North is to the top and East is to the left.  The image size is 7.65 by 8.40 arc-minutes. \label{fig-mosaic}}
\end{figure}

\begin{figure}
\plotone{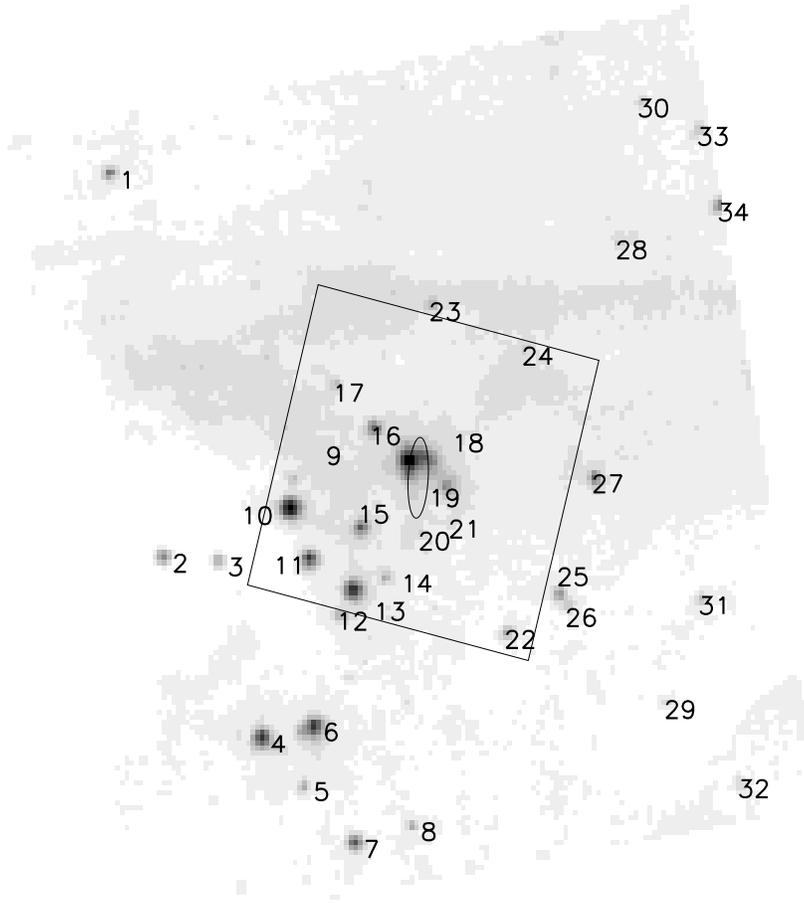}
\caption{Same as Figure 1 except with the identification labels for the sources identified.  The region marked with the solid line shows the coverage of the CVF data.\label{fig-m2}}
\end{figure}

\begin{figure}
\plottwo{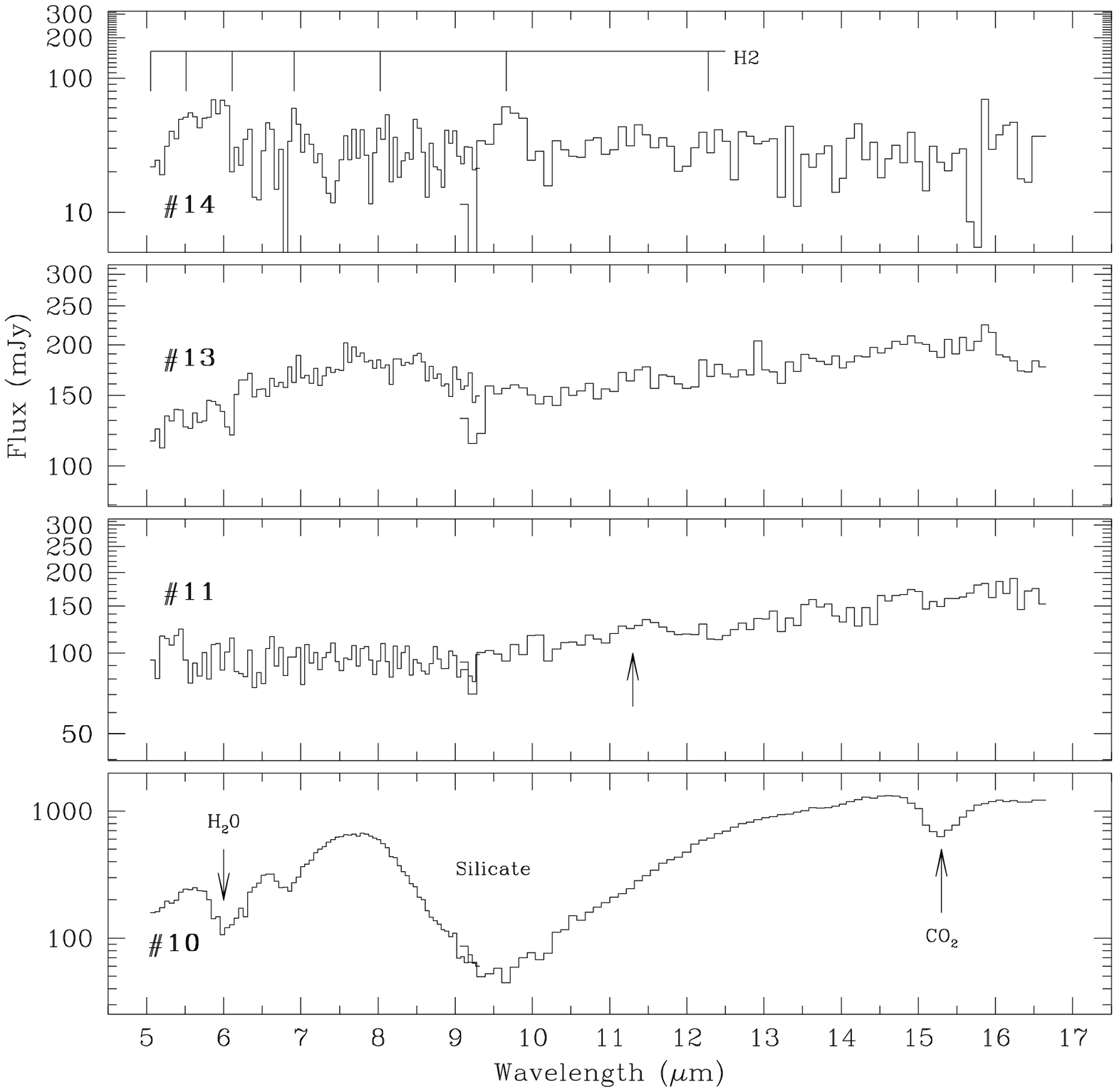}{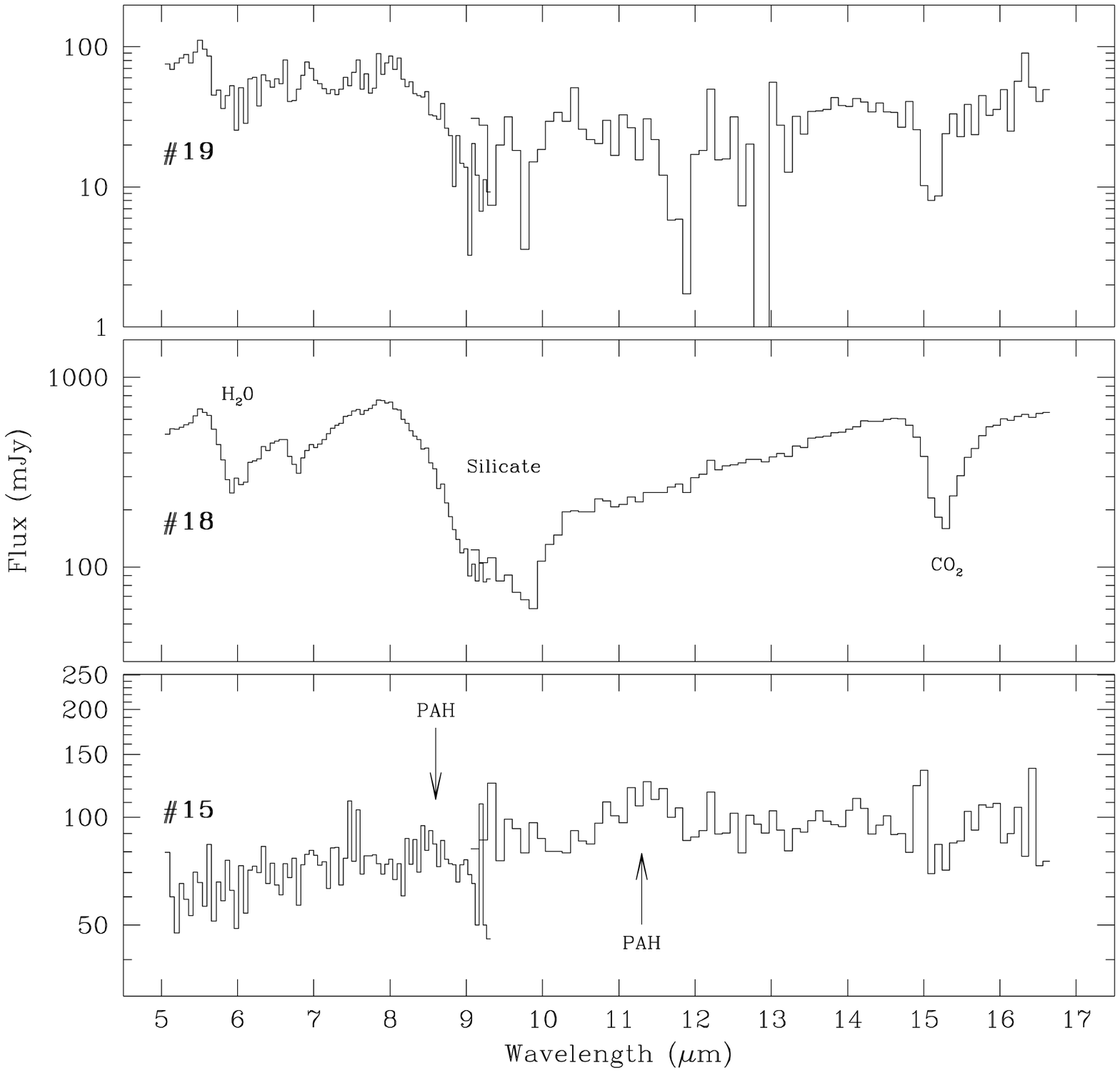}
\caption{The low-resolution spectra of 7 sources scanned with the CVF.  Each object is identified in the left-hand side of the panel. The prominent spectral features are identified. These spectra are consistent with those seen for the youngest proto-stars in star-forming regions. \label{fig-cvfspec}}
\end{figure}

\begin{figure}
\plotone{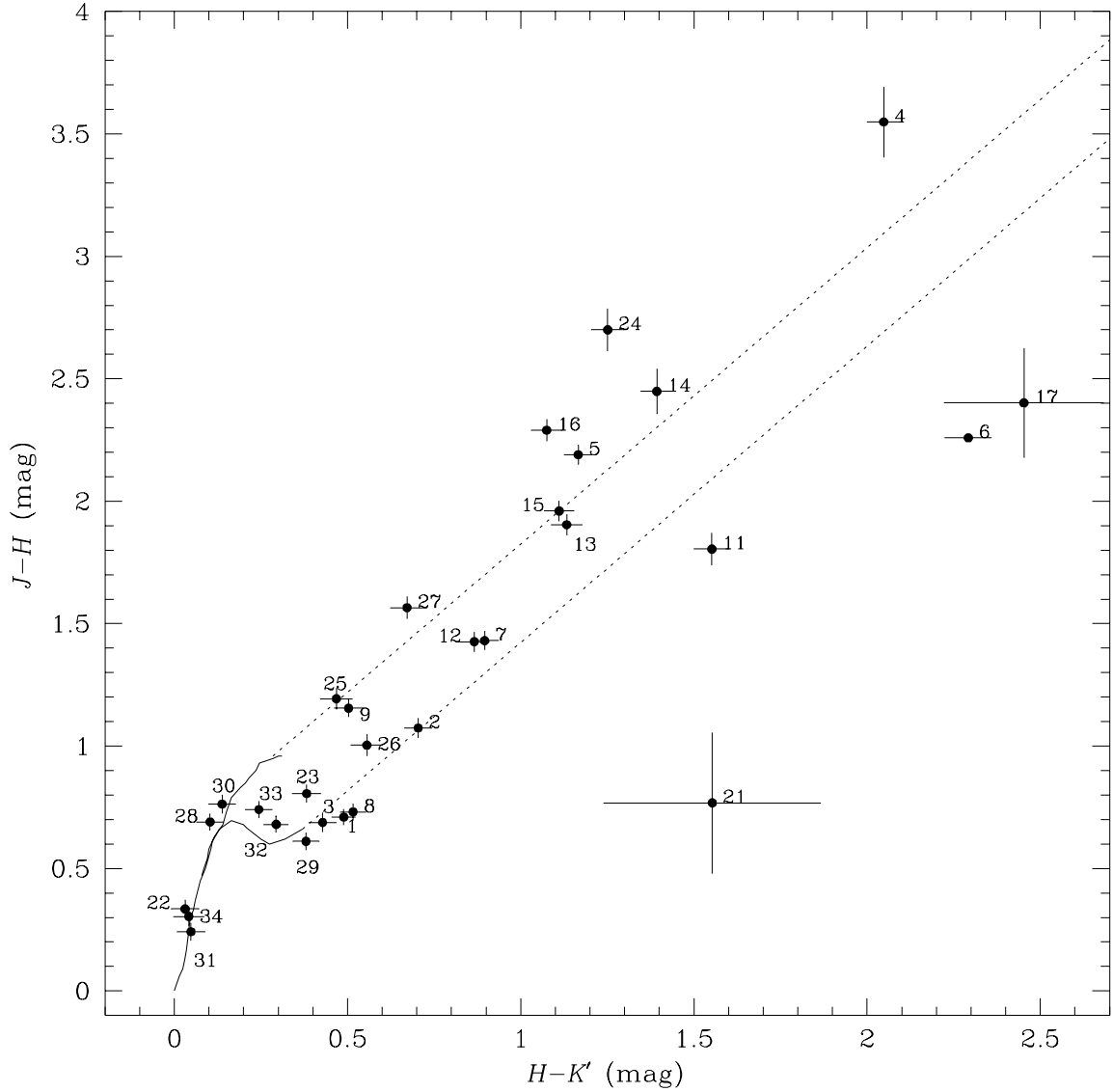}
\caption{The {\protect \jmh \vs \hmk} color-color relationship for all ISOCAM sources also detected by 2MASS.  The solid lines show the location of normal dwarfs and giants on this diagram.  The dotted lines show the reddening vectors.  Most of the L 1641-N sources appear reddened similar to other star-forming regions.  Four sources show evidence for "excess", presumably circumstellar, emission.\label{fig-nirclrs}}
\end{figure}

\begin{figure} 
\plotone{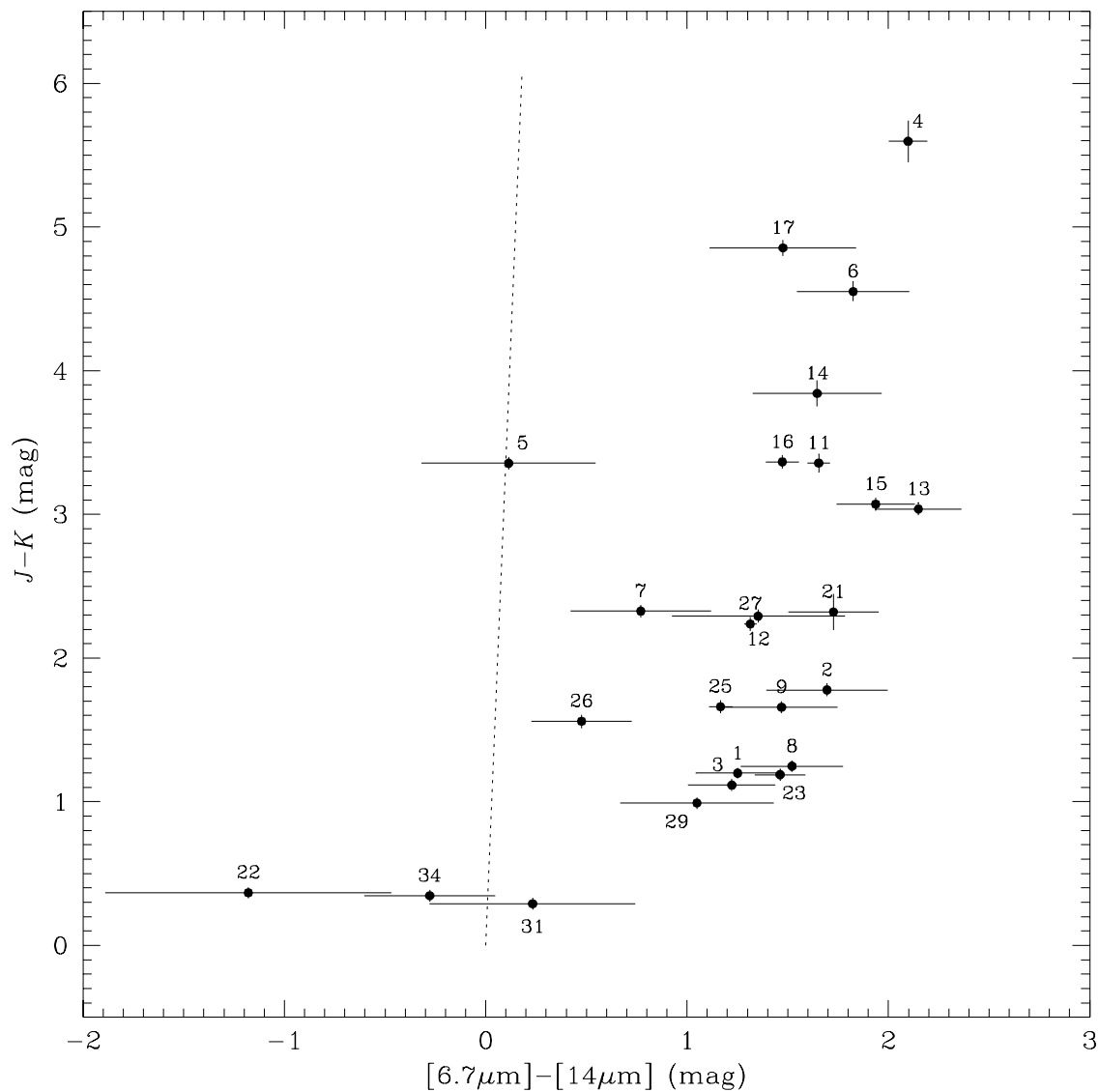}
\caption{The {\protect \jmk \vs $\isotwothree$} color-color relationship of all sources detected in all 4 filters.  The lines have the same meaning here as in Figure~\ref{fig-nirclrs}. All but four source seen show evidence for excess emission.  Thus, at mid-IR wavelengths, at least 85\% of the sources show excess emission, consistent with Persi et al. (2000) who find 75\% of the sources in the Chamaeleon I cloud show excess emission at mid-IR wavelengths.\label{fig-midclrs}}
\end{figure}

\end{document}